\documentclass[%
 reprint,
 superscriptaddress,
footinbib,
 amsmath,amssymb,
 aps,
prx,
floatfix,
]{revtex4-2}
\usepackage{graphicx}
\usepackage{dcolumn}
\usepackage{bm}
\usepackage{siunitx}
\usepackage[T1]{fontenc}
\usepackage{orcidlink}
\usepackage[final]{pdfpages}
\setboolean{@twoside}{false}
\makeatletter
\AtBeginDocument{\let\LS@rot\@undefined}

\usepackage{multirow}
\usepackage[columnwise]{lineno}
\usepackage{algorithm}[compatible]
\usepackage[noend]{algpseudocode} 
\usepackage{nicefrac}
\usepackage[export]{adjustbox}
\usepackage{rotating}

{\footnotetext{Present address: Photonic, Inc., Vancouver, Canada.}}
\usepackage{color}

\usepackage{hyperref}

\begin{document}
\preprint{}
\title{Bootstrapping, autonomous testing, and initialization system for Si/Si$_x$Ge$_{1-x}$ multi-quantum-dot devices}

\author{Tyler J. Kovach\orcidlink{0009-0007-0807-7300}}
\thanks{Contact author: \href{tkovach@wisc.edu}{tkovach@wisc.edu}}
\affiliation{Department of Physics, University of Wisconsin-Madison, Madison, Wisconsin 53706, USA}

\author{Daniel Schug\orcidlink{0009-0001-3758-501X}}
\affiliation{Department of Computer Science, University of Maryland, College Park, Maryland 20742, USA}

\author{M. A. Wolfe}
\affiliation{Department of Physics, University of Wisconsin-Madison, Madison, Wisconsin 53706, USA}

\author{E. R. MacQuarrie$^\S$\orcidlink{0000-0003-3097-5549}}
\affiliation{Department of Physics, University of Wisconsin-Madison, Madison, Wisconsin 53706, USA}

\author{Patrick J. Walsh\orcidlink{0009-0004-6683-5817}}
\affiliation{Department of Physics, University of Wisconsin-Madison, Madison, Wisconsin 53706, USA}

\author{Owen M. Eskandari\orcidlink{0009-0009-3197-9673}}
\affiliation{Department of Physics, University of Wisconsin-Madison, Madison, Wisconsin 53706, USA}

\author{Jared Benson\orcidlink{0009-0009-1673-5259}}
\affiliation{Department of Physics, University of Wisconsin-Madison, Madison, Wisconsin 53706, USA}

\author{Mark Friesen\orcidlink{0000-0003-2878-2844}}
\affiliation{Department of Physics, University of Wisconsin-Madison, Madison, Wisconsin 53706, USA}

\author{M. A. Eriksson\orcidlink{0000-0002-3130-9735}}
\affiliation{Department of Physics, University of Wisconsin-Madison, Madison, Wisconsin 53706, USA}

\author{Justyna P. Zwolak\orcidlink{0000-0002-2286-3208}}
\thanks{Contact author: \href{jpzwolak@nist.gov}{jpzwolak@nist.gov}}
\affiliation{National Institute of Standards and Technology, Gaithersburg, Maryland 20899, USA}
\affiliation{Joint Center for Quantum Information and Computer Science,
University of Maryland, College Park, MD 20742, USA}
\affiliation{Department of Physics, University of Maryland, College Park, Maryland 20742, USA}

\date{\today}
\begin{abstract}
Semiconductor quantum dot (QD) devices have become central to advancements in spin-based quantum computing. 
However, the increasing complexity of modern QD devices makes calibration and control---particularly at elevated temperatures---a bottleneck to progress, highlighting the need for robust and scalable autonomous solutions.
A major hurdle arises from trapped charges within the oxide layers, which induce random offset voltage shifts on gate electrodes, with a standard deviation of approximately 83~\si{\milli\volt} of variation within state-of-the-art present-day devices.
Efficient characterization and tuning of large arrays of QD qubits depend on choices of automated protocols. 
Here, we introduce a physically intuitive framework for a bootstrapping, autonomous testing, and initialization system (BATIS) designed to streamline QD device evaluation and calibration.
BATIS navigates high-dimensional gate voltage spaces, automating essential steps such as leakage testing, formation of all current channels, and gate characterization in the presence of trapped charges.
For forming the current channels, BATIS follows a non-standard approach that requires a single set of measurements regardless of the number of channels.
Demonstrated at $1.3$~\si{\kelvin} on a quad-QD Si/Si$_x$Ge$_{1-x}$ device, BATIS eliminates the need for deep cryogenic environments during initial device diagnostics, significantly enhancing scalability and reducing setup times. 
By requiring only minimal prior knowledge of the device architecture, BATIS represents a platform-agnostic solution, adaptable to various QD systems, which bridges a critical gap in QD autotuning.
\end{abstract}

\maketitle
\section{Introduction}
Quantum dot (QD) qubit devices are commonly formed by electrostatically manipulating the potential energy landscape of a two-dimensional (2D) electron gas (2DEG) or hole gas (2DHG).
This procedure produces interleaved regions of accumulated and depleted charge density that serve as QD qubits, charge sensors, and charge carrier reservoirs~\cite{Burkard23-SSQ, Wiel02-DQD, Loss98-QCD, Hanson07-SQD, Zwanenburg13-SQE}.
In recent years, the number of QDs in such devices has risen rapidly~\cite{Ha22-FDQ, Wang2024, George24-SAM}.
In the near term, it will no longer be possible to tune up QD qubit arrays manually.
The field of autotuning has arisen to address this challenge~\cite{Zwolak21-AAQ, Zwolak23-DNC}.

Modern semiconductor QD qubit devices have two critical features that enhance their performance while simultaneously making autotuning challenging and essential.  
First, the active regions of the device---the qubit QDs, charge sensors, and reservoirs---contain no doping.
All charge carriers---electrons or holes---are induced by gate voltages, and no current can flow until gate voltages are applied and carriers are accumulated.
At the same time, the charges trapped within the gate-oxide heterostructure of the QD devices make it impossible to know \textit{a priori} the precise turn-on voltages, i.e., the gate voltage threshold at which carrier accumulation begins.  
The device drift and hysteresis (i.e., a shift of the current curve that depends on the direction of the sweep) introduce additional variability in voltage configuration not only between cooldowns but even between successive initialization steps.
Second, QD device chips have multiple current channels that often serve different purposes.
Some host an array of QD qubits, while others provide a home for charge sensors.  
These channels are not separated until pinch-off voltages, i.e., the gate voltage threshold at which carrier accumulation is eliminated, are determined~\footnote{Mathematically, the pinch-off and the turn-on voltage are indistinguishable for a given current curve.
However, when measuring a physical gate, a hysteresis can be observed.}. 
The size of the QD introduces additional challenges: the accumulation and depletion of carriers must occur in alternating regions separated by only tens of nanometers. 
While spatially separated, the channels must remain coupled to ensure proper operation of proximal charge sensors on nearby QDs.

In this work, we introduce and demonstrate an intuitive bootstrapping, autonomous testing, and initialization system (BATIS) that can automatically manipulate gate voltages to ensure proper formation of all current channels.
BATIS takes as input the device gate layout and a high-level understanding of the desired operation, including (critically) where current channels should be continuous and which gates should divide one channel from another.
The output of BATIS is a comprehensive device characterization, specifying the gate voltage settings needed for QD chip operation, with clearly delineated current channels and identified pinch-off voltages.

Previous work on autotuning has shown that correctly executed bootstrapping of QD devices sets the stage for the more specialized tuning phases~\cite{Zwolak21-AAQ}. 
For example, setting the device topology has been demonstrated for undoped accumulation-mode QD devices~\cite{Zwolak20-AQD, Zwolak21-RBI}.
Automated calibration of the charge state has been demonstrated for a modulation-doped QD device~\cite{Durrer19-ATQ}, an undoped silicon-metal-oxide-semiconductor QD device~\cite{Lapointe-Major19-ATQ}, and undoped accumulation-mode QD devices~\cite{Ziegler22-TAR}.

Methods for automatically detecting unintentional, spurious QDs have also been proposed~\cite{Ziegler23-AEC}.
More recently, virtual-gate control of the electrostatic potential landscape confining holes within a 2D QD array has been demonstrated~\cite{Rao24-MAViS}.

Automated tuning of QD devices from a bootstrapping has been demonstrated for modulation-doped QD devices with predefined current channels~\cite{Baart16-CAT, Darulova19-ATQ, Severin21-CAT} as well as for silicon in field-effect transistors and nanowires, which by design have a single current channel~\cite{Severin21-CAT}.
However, all these demonstrations assumed prior knowledge of the device and a non-negligible device-specific preconfiguration.
The first fully autonomous bootstrapping algorithm, intended for initiation on a pristine modulation-doped QD device, was introduced in Ref.~\cite{Zubchenko24-ABQ}; however, the proposed algorithm is platform specific and assumes the QD device to be functional.
Autonomous characterization and bootstrapping for accumulation-mode QD devices have never been attempted to our knowledge.

BATIS addresses a key challenge in bootstrapping the QD devices: the diversity of device architectures and control layouts. 
All initialization tools proposed to date rely on rigid measurement sequences tailored to specific device designs, limiting their applicability across platforms. 
BATIS, in contrast, is designed to be platform agnostic and thus can operate across a wide range of QD devices by dynamically adapting its procedures.
While the procedural flow of BATIS is well-defined at a high level, the \textit{executable flow}---the actual sequence of measurements and operations---is automatically adjusted in real time on the basis of the device’s configuration and control wiring, as specified in a flexible configuration file. 
In other words, throughout execution, BATIS adjusts which steps to take and what outcomes to expect on the basis of the device's observed behavior.
As tuning progresses, BATIS further tailors subsequent measurements to match the discovered characteristics of the system, ensuring efficient and device-specific calibration without manual intervention.
 
Importantly, the sample used to demonstrate BATIS is kept at around 1.3~\si{\kelvin}, which is about 10 times warmer than other typical devices~\cite{Borsoi22-QCA, Park24-SSP}.
High-temperature operation is critical for practical quantum computer implementations~\cite{Vandersypen17-ISQ}.
However, higher temperatures broaden key measurement features, such as Coulomb blockade peaks, necessitating careful fitting procedures to extract the desired parameters~\cite{PhysRevLett.65.771}.
The industry has demonstrated promising efforts in characterizing the performance of industry-manufactured spin qubit devices at 1.6~\si{\kelvin} using the cryogenic 300-\si{\milli\meter} wafer prober~\cite{Kruijf-23-CPS, Neyens24-PQW}.
Autotuning of a single-electron transistor charge sensor in Si at 1.5~\si{\kelvin} has also been demonstrated~\cite{Paurevic25-QCS}.

We demonstrate that BATIS performs reliably at temperatures above 1.0~\si{\kelvin} and in the presence of trapped charges. 
It autonomously adapts to the specific QD layout and carries out critical initial tune-up stages, including identifying fabrication defects and characterizing both local and global current flow. 
These capabilities directly address emerging bottlenecks in scaling quantum semiconductor chips, making BATIS a powerful tool for enabling large-scale qubit implementations.

This paper is organized as follows: In Sec.~\ref{sec:problem},  we provide the motivation for the necessity of autotuning from the perspective of trapped charges.
The bootstrap tuning procedure is described in detail in Sec.~\ref{sec:tuning_procedure}.
The experimental demonstration of BATIS to fully autonomously tune up an accumulation-mode Si/Si$_x$Ge$_{1-x}$ QD device at 1.3~\si{\kelvin} is described in Sec.~\ref {sec:exp_run}.
We conclude with a discussion of future directions in Sec.~\ref{sec:conclusion}.

\section{Trapped Charge: A Barrier to Robust Control}
\label{sec:problem}
A primary challenge in automating current channel formation in QD devices stems from the presence of trapped charges within the QD array, which introduce substantial variability in the required gate voltages.
``Charge trapping'' refers to the immobilization of carriers at unknown locations due to, for example, structural imperfections and defects on the surfaces and at interfaces of the heterostructure hosting the QDs.
Stray charges can accumulate in any region of the heterostructure, including at buried interfaces, in the oxide layer at the top of the stack, and in particular, at the semiconductor-oxide interface, where there is an extremely high density of interface states~\cite{SzeBook}.

Trapped charges in QD arrays present multiple challenges that degrade device performance~\cite{Massai24-ITC}. 
First, they reduce the effective carrier concentration, which impedes transport and reduces mobility. 
In addition, trapped charges introduce temporal fluctuations in the local electric field, resulting in increased charge noise that compromises the stability and coherence of quantum states within the QDs. 
Over time, charge trapping can also lead to electrostatic disorder at the interface between the QD array and surrounding materials, further complicating device tunability and limiting the precision required for quantum control.

While trapped charges are largely locked into place as the device is cooled down from room temperature, they can nevertheless slowly drift over time, even at millikelvin temperatures~\cite{Rudolph2019}. 
Moreover, these processes are not well controlled, leading to nonuniform charge accumulation that can vary from one cooldown to the next.
As a result, significant and nonuniform voltage shifts may be required to offset the effects of trapped charges and achieve uniform gate operation across a device, even for nominally identical gate electrodes within a single device.

Such variability is undesirable for large-scale applications, and several techniques have been developed to mitigate this problem~\cite{Meyer2023}.
One such approach involves a simple ``reset'' of the trapped charges by optically illuminating the device \textit{in situ} using a laser diode~\cite{Wolfe24-COI}.
Illumination temporarily excites electrons above the band gap of the host semiconductor, enabling the redistribution of immobile, unwanted trapped charges that otherwise remain fixed at low temperatures.
Unfortunately, this procedure is not guaranteed to improve device performance and often needs to be repeated several times through trial and error, with different applied gate voltages.

\begin{figure}[t]
    \centering
    \includegraphics[width=0.98\columnwidth]{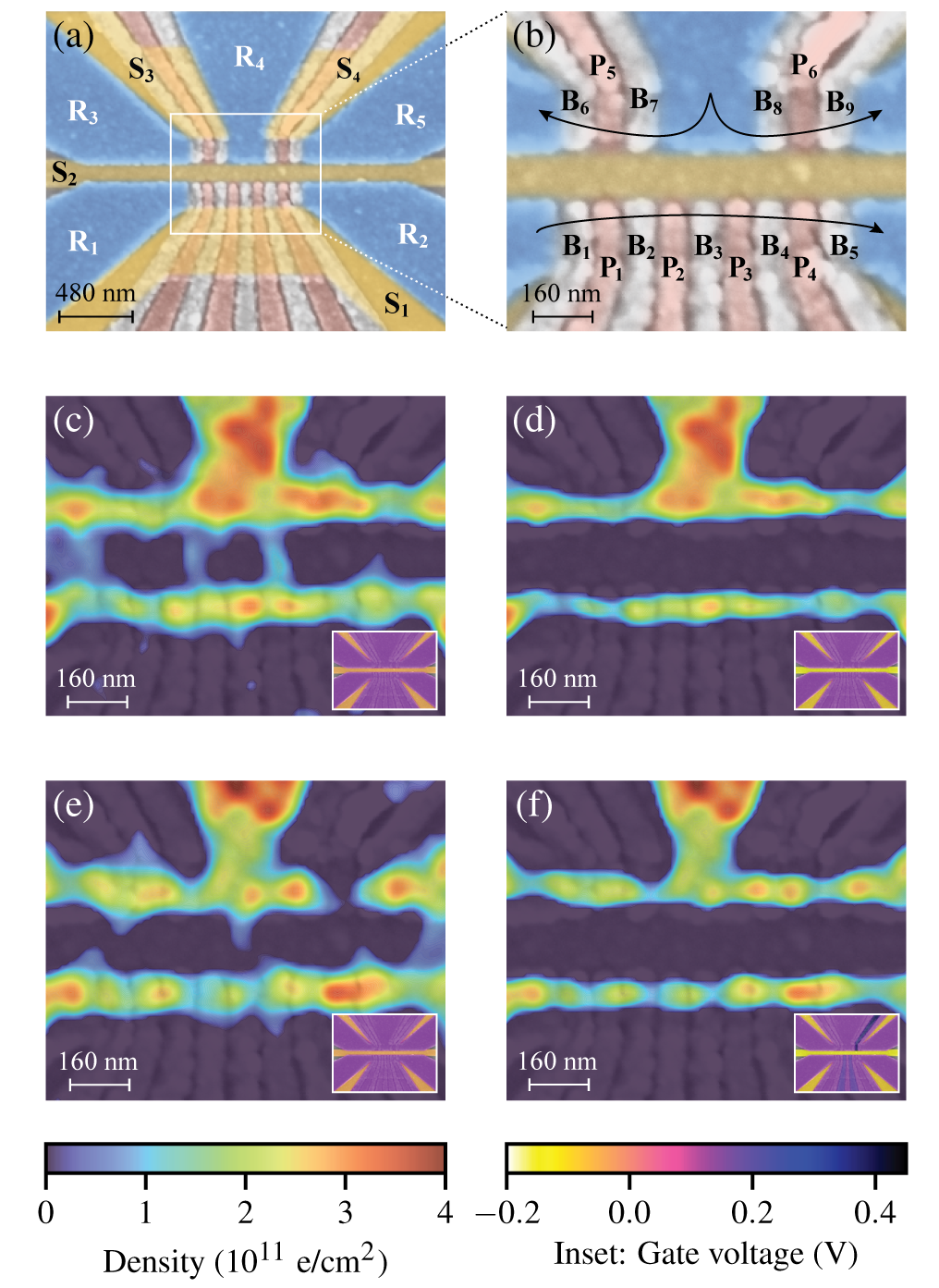}
    \caption{
    (a) False-color scanning electron microscopy image of a device nominally identical to one used in this work. 
    The color indicates the gate's function as a screening (S$_1$-S$_4$; yellow), reservoir (R$_1$-R$_5$; blue), barrier (B$_1$-B$_9$; gray), or plunger (P$_1$-P$_6$; red) gate.
    (b) Higher-magnification image with the desired current paths shown by black arrows and the individual barrier and plunger gates labeled.
    (c-f) Self-consistent Schr\"odinger-Poisson simulations (density plots), showing typical effects of trapped charge on 1D channel formation in a 2DEG.
    The simulation results are overlaid on the scanning electron microscopy image.
    (c) Trapped charge can lead to current paths crossing the central screening gate; an autotuner must detect and correct for this problem, as shown in (d).
    (e) Trapped charge can also lead to breaks in the current paths, another possible error that an autotuner must detect and correct for, as shown in (f).
    The insets in (c)--(f) show the gate voltages necessary to cause the charge densities.
    For all simulations, interface trapped charge is chosen, as described in the main text and in the Supplemental Material~\cite{supp}.
    }
    \label{fig:physical_problem}
\end{figure}

The effect of trapped charges can be illustrated by simulating channel formation in a typical QD device.
Figures~\ref{fig:physical_problem}(a-b) show a gate-defined, accumulation-mode, undoped QD device with multiple current channels, six designated regions for forming QDs (under the six \textit{plunger} gates, shown in red), and \textit{reservoir} gates (shown in blue) to connect these regions to Ohmic contacts for current flow.
This device exemplifies a commonly used fabrication strategy known as overlapping gates architecture~\cite{Zajac16-SGA, Dodson20-FQD}.

The \textit{screening} gates (shown in yellow) form the bottommost layer in the three-layer gate architecture, positioned closest to the gate oxide. 
Their primary role is to screen electric fields, thereby preventing the unintentional formation of 2DEG or 2DHG regions and electrically isolating the intended transport channels.
The second layer includes the reservoir and plunger gates.
The reservoirs enable the flow of electrons from the Ohmic contacts toward the center of the device, where QDs form under electrostatic confinement.
Each plunger gate primarily controls the chemical potential of the QD below it.
The final layer consists of \textit{barrier} gates (shown in gray), which control the tunnel couplings between QDs---both qubit QDs and charge-sensing QDs---and reservoirs.
By careful adjustment of the voltages applied to each gate, the QD qubit chip with charge sensors is made functional~\cite{Wiel02-DQD, Loss98-QCD, Hanson07-SQD, Zwanenburg13-SQE}.

To model the effects of trapped charge, we introduce randomly varying charge densities on a tiled grid array at the oxide-semiconductor interface, with the density of each tile selected at random from a normal distribution, with the standard deviation corresponding to a shift magnitude $\Delta V=83$~\si{\milli\volt}, consistent with experimental observations for silicon QDs~\cite{Neyens24-PQW}.
Each tile is $40$~\si{\nano\meter} by $50$~\si{\nano\meter}, ensuring that all gates encompass multiple regions of randomly selected charge densities.

\begin{figure*}[ht]
    \centering
    \includegraphics[width=0.97\textwidth]{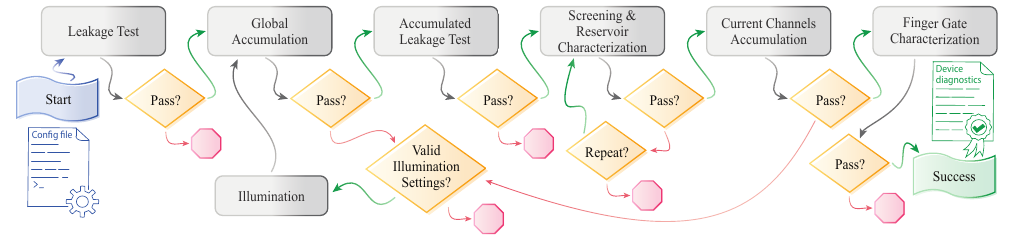}
    \caption{BATIS algorithm.
    The flowchart representation of the procedural flow of BATIS.  
    Beyond manipulating gate voltages, BATIS can also change the device response to gate voltages \textit{in situ} by illuminating the device. 
    Active steps in autotuning are shown by rounded gray rectangles, conditional logic is shown by yellow diamonds, and failure determinations are shown by red stop signs.
    The green and red arrows indicate whether a given step passes or fails, respectively.
    }
    \label{fig:algorithm}
\end{figure*}

In the case where no random trapped charge is included, the ideal charge densities in the quantum well form coherent channels half way between the screening gates, as indicated by the arrows in Fig.~\ref{fig:physical_problem}(b).
In contrast, Figs.~\ref{fig:physical_problem}(c) and \ref{fig:physical_problem}(e) show results obtained when randomly distributed trapped charges are included, resulting in current channels that flow beneath screening gates and pinched channels, respectively.
Both problems can be resolved by careful adjustment of the various gate voltages to recover the coherent channels shown in Figs.~\ref{fig:physical_problem}(d) and \ref{fig:physical_problem}(f), respectively.

While misdirected and pinched channels are easy to identify in simulations, they are much more difficult to diagnose in experiments.
The issue of current flowing beneath the screening gate and unintentionally connecting multiple channels, as illustrated in Fig.~\ref{fig:physical_problem}(c), is intrinsic to 2D gate-defined systems and does not exist in one-dimensional (1D) nanowire and nanotube devices~\cite{Wang22-UQD, Tormo22-NDD}; therefore, it is of particular interest in this work.
The recovery of misdirected current to form the desired coherent channels requires the simultaneous tuning of multiple gates.
The protocols developed in this work are specially designed to address and characterize such 2D challenges.

\begin{figure*}[ht]
    \centering
    \includegraphics[width=0.98\textwidth]{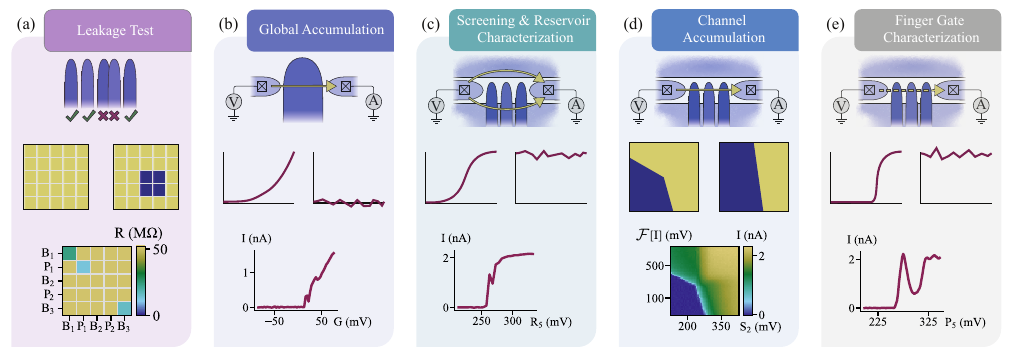}
    \caption{Stages of the tuning procedure.
    The first row in (a)--(d) explains the measurements.
    The second row shows idealized cartoon representations of the key data for each stage of the tuning process, with good and bad data shown on the left and the right, respectively. 
    The third row provides examples of experimental data.  
    The yellow arrows in the cartoons indicate the ideal current paths.
    (a) In the leakage test, no gate should have a low resistance to any other gate or cryostat ground (Ohmic contacts are included in the test for the nonaccumulated leakage test).
    The good cartoon matrix shows high-resistance paths on all connections.
    The bad cartoon matrix shows two gates shorted to each other.
    The realistic matrix shows three gates shorted to ground.  
    Devices with shorted gates are typically unusable. 
    (b) When accumulating the 2DEG in the device globally, the current is expected to rise above zero with the increase of gate voltages, as seen in the good cartoon plot.
    The bad cartoon plot shows a failure to turn on.
    A sccessful turn-on is observed in the experimental data, with G representing $\{{\rm S}_1, {\rm S}_3, {\rm S}_4, {\rm R}_1, {\rm R}2, {\rm R}3, {\rm R}4, {\rm R}5\}$ set of gates. 
    (c) For pinch-off, it is desirable for the current to drop to the noise floor as the gate voltage is lowered. 
    The bad cartoon demonstrates a failure of the screening or reservoir gate to pinch off.
    The experimental data show some small oscillations due to, for example, accidental Coulomb blockade during pinch-off.
    (d) For channel accumulation, rather than relying on measuring pairs of screening gates, we sweep a single screening gate against all the finger gates corresponding to that current channel. 
    The bad cartoon measurement indicates that at least one finger gate is too far below the threshold to accumulate.
    (e) Each plunger and barrier gate must pinch off individually.
    The bad cartoon shows the finger gate failing to pinch off.
    In the experimental data, Coulomb blockade oscillations are often observed and are particularly prominent during pinch-off at this tuning step.
    }
    \label{fig:physics_steps}
\end{figure*}

\section{BATIS: Design and implementation}
\label{sec:tuning_procedure}
BATIS is designed to automate the testing, characterization, and initialization of all gates, as well as the formation of all necessary 1D current channels for linear QD arrays.
The procedural flow of the single-button-press algorithm, depicted in Fig.~\ref{fig:algorithm}, is rooted in the underlying device physics, making it both interpretable and robust, while also enabling adaptability across devices and scalability to arbitrary-length 1D arrays.

BATIS takes as input two configuration files describing the device design and certain generic specifications of the device and instruments.
These files include detailed descriptions of all gates, the desired current channels, and the QD device connectivity. 
Additionally, they define a wire map of the experimental setup and a set of variables associated with each stage of the tuning process.
These variables include instrumentation constraints, which prevent the algorithm from exploring unsupported operating ranges, and device safety limits, which protect the hardware from excessive or damaging voltages. 
Using this information, BATIS performs comprehensive diagnostics and characterization of each gate and automatically determines the appropriate configuration for the next stage of the tuning sequence, as depicted in Fig.~\ref{fig:algorithm}.

The configuration files establish a link between abstract descriptions of device components and the operational language used within BATIS. 
For instance, the functionality of each gate is automatically inferred from the user-provided device architecture, enabling BATIS to construct a gate lookup table. 
When combined with the wire map, this allows BATIS to automatically identify and perform the appropriate specialized measurements by linking the logical gate definitions to the physical measurement instruments. 
This abstraction layer ensures that BATIS remains agnostic to both the specific device layout and the instrumentation used.

The algorithm follows conditional logic, which is visually represented by boxes and arrows of different colors in Fig.~\ref{fig:algorithm}.
Each of the six stages in BATIS, indicated in Fig.~\ref{fig:algorithm} with rounded gray rectangles, involves at least one optimization of a device measurement or manipulation followed by data analysis to extract characteristics of a selected subset of gates.
The measurements typical of each stage are detailed in Fig.~\ref{fig:physics_steps}, with the first row in each panel explaining the measurement, and the second and third rows showing a cartoon representation of the data (in each case, showing success on the left and failure on the right) and an example of measured experimental data for each step, respectively.

The procedural steps in BATIS are indicated by gray arrows. 
Passing the functionality test for the current stage, marked with yellow diamonds in Fig.~\ref{fig:algorithm}, initiates the next stage, as indicated by the green arrows.
Failing a test, indicated by red arrows, results in repeated measurements, device optical illumination to adjust the trapped charges, or bootstrapping termination, depending on the issues revealed by BATIS and the conclusions it draws on the basis of the device physics.
In the present implementation, illumination is treated as a rare, high-impact recovery step rather than a continuously tunable control knob. 
Each illumination heats the cryostat and, therefore, only a small number of illumination attempts are feasible per cooldown. 
Within this very small, safety-constrained action space, a simple, physics-motivated decision tree provides a transparent and effectively sample-optimal policy for illumination, establishing a baseline against which more sophisticated controllers can later be benchmarked.

When the procedures have been completed, BATIS produces a device diagnostic file that stores all information about each gate learned during the tuning process. 
When executed successfully, it verifies that all gates are functioning as intended and that isolated regions can be formed in well-defined 1D channels beneath the designated plunger gates.
It also preconfigures the QD device for further use.
In the following subsections, we explain each state of the algorithm.

\subsection{First leakage test}
\label{sec:first_leakage}
When initiated, BATIS first checks the device for leakage to the fridge ground through the oxide or any other possible means in the instrumentation setup.
Leakage can be caused, for example, by defects arising during fabrication or packaging. 

The first leakage measurement happens before the electron accumulation, during the \textit{leakage test} stage; see Fig.~\ref{fig:algorithm}. 
The resistances of all the connections are measured with respect to each other and the fridge ground, and they are compiled into a \textit{leakage matrix}, as exemplified in the second and third rows in Fig.~\ref{fig:physics_steps}(a). 
Leakage manifests itself as off-colored cells in the matrix, as shown in the right panel in the second row and in the third row in Fig.~\ref{fig:physics_steps}(a).
The desired result is a high resistance between all the connections.
Should leakage be detected, BATIS terminates and flags the device as unusable.

\subsection{Global accumulation}
\label{sec:global_accumulation}
For accumulation-mode QD devices, no 2DEG is present at zero applied gate voltage. 
Application of a nonzero positive bias lowers the conduction band minimum below the Fermi level, resulting in accumulation of a 2DEG in the quantum well near the Si/Si$_x$Ge$_{1-x}$ interface. 
The goal of the \textit{global accumulation} stage is to determine the characteristic turn-on voltage $V_\mathcal{P}$ where enough of the 2DEG is accumulated that a significant amount of current begins to flow between the Ohmic contacts, effectively turning the device on; see Fig.~\ref{fig:physics_steps}(b). 
Since BATIS assumes no \textit{a priori} knowledge of the device response, $V_\mathcal{P}$ is determined through an iterative approach, with $V_{\rm max}$ increased incrementally until BATIS is confident that $V_\mathcal{P}$ is observed in the measured window. 

Idealized successful and failing measurements for this stage are visualized in the left and right panels, respectively, in the second row in Fig.~\ref{fig:physics_steps}(b).
The turn-on process in real QD devices is not always straightforward, as even fully functional devices may fail to turn on correctly during initial testing. 
If the algorithm encounters a problem during global turn-on, it has a recovery pathway through optional \textit{illumination} via 780~\si{\nano\meter} light to adjust the microscopic configuration, magnitude, and sign of trapped charge, which can dramatically alter the device's response to gate voltages~\cite{Wolfe24-COI}; see Fig.~\ref{fig:algorithm}.

After illumination, the global accumulation stage is reinitiated.
If the preference is not to use illumination, BATIS allows skipping of this step and proceeding directly to repeating the global turn-on stage.
Whether or not illumination should be used is an option that is predefined in the configuration file.

Determination of the turn-on voltages $V_\mathcal{P}$ relies on fitting to a phenomenological function that mimics the expected behavior of the signal.
Before fitting, the raw data are preprocessed to clean the signal.
The preprocessing involves cropping the data to avoid nonlinearities at the extremes of the data-acquisition electronics, followed by smoothing to remove noise from the signal and normalization. 
We then parameterize the $I$-$V$ response of the device using 
\begin{equation}\label{eq:phen_fit}
    I \propto \left[1+\exp\left(-\frac{V-V_0}{\delta V}\right)\right]^{-1},
\end{equation}
where $V_0$ is the center of the step function and $\nicefrac{1}{4\delta V}$ is the slope at $V_0$.
This function mimics the expected exponential behavior of the Fermi-Dirac distribution of electron energies with respect to gate voltages~\cite{Dirac26-OQM, Zannoni99-OIG, Bardeen47-SSC}.
Similarly to the $I$-$V$ characteristics of metal–oxide–semiconductor field-effect transistors, an expected saturation will occur, leading to a roughly sigmoidal shape \cite{Dacey53-FET}.

The final turn-on voltage is recorded as
\begin{equation}\label{eq:vp}
    V_{\mathcal{P}} = V_0 + 8v\,\delta V,
\end{equation}
where $v=-0.5$ is a fixed constant specifying how much the gate needs to be adjusted to turn on in the unit of $\delta V$~\footnote{The parameter $v$ is defined as part of the configuration file and can be adjusted if needed.}.
The safe maximum allowed voltage applied to each gate, V$_{\rm max}$, is then updated to better match the saturation of the curve, such that it is less than or equal to the original.
This voltage is later used as the starting point for the characterization of the screening and reservoir pinch-offs.

\subsection{Accumulated leakage test}
\label{sec:accumulated_leakage_test}
With the global accumulation confirmed, BATIS proceeds to the accumulated leakage test.
This step proceeds analogously to the first leakage check, except that it excludes the Ohmic connections.
This is because this leakage check is performed when the 2DEG in the device is accumulated, with the Ohmic connections intentionally shorted to each other.
If this stage fails, it indicates that a gate is shorted to the 2DEG flowing beneath it, which is an unrecoverable fabrication defect.
Should that occur, BATIS terminates.

\subsection{Screening and reservoir characterization}
\label{sec:screening_and_reservoir_characterization}
The global accumulation voltage does not provide gate-specific information.
To learn that information, BATIS performs a series of individual gate characterizations. 
The first of these is performed at the \textit{screening and reservoir characterization} stage; see Fig.~\ref{fig:algorithm}. 

With the 2DEG globally accumulated, BATIS depletes the voltages on individual gates until the current on the corresponding channel is pinched off, as displayed in the left panel in the second row in Fig.~\ref{fig:physics_steps}(c). 
BATIS uses these pinch-off curves to determine the operating points for reservoir gates, ensuring proper 2DEG accumulation, and to determine the operating bounds for the screening gates. 

The derivative of the $I$-$V$ response signals for screening and reservoir gate pinch-offs are preprocessed in a similar fashion as described in Sec.~\ref{sec:global_accumulation}.
The pinch-off, guaranteed isolation, and operating voltages are defined following Eq.~(\ref{eq:vp}), with $v=-0.5$, $v=-1.0$, and $v=0.5$, respectively.
The pinch-off occurs at roughly the inflection of the signal $V_\mathcal{P}$, the guaranteed isolation V$_\mathcal{I}$ happens left of the pinch-off, and the operating voltages V$_\mathcal{O}$ occur to the right of the pinch-off,  typically on the slope. 
The operating voltage for the central screening gate is handled separately, as it has to satisfy a secondary constraint: all current channels must be formed without merging.
By default, $v=-0.75$.
If at the subsequent stage a failure to form 1D current channels is observed twice, BATIS adjusts this parameter in increments defined as $v_{\tau}=(-1)^{p}\,0.02$, where $p=\{0,1\}$ is chosen at random. 

Typical unsatisfactory operational modes for this stage include too much depletion or accumulation; see the right panel in the second row in Fig.~\ref{fig:physics_steps}(c).
If the 2DEG along the channel below a gate is extraordinarily depleted, no voltage can be applied to the gate to accumulate it.
On the other hand, if the channel is too accumulated, it cannot be pinched off. 
However, both types of issue would have been detected at the global accumulation stage, prompting the initiation of device illumination.
Thus, at this stage, the constant current signal is caused by either an instrument failure or a broken gate. 
To verify this, BATIS conducts a repeated measurement. 
If the error is observed a second time, this indicates an issue with the gate, and the algorithm terminates.

\subsection{Current channel accumulation}
\label{sec:channel_accumulation}
The 1D current channels are formed by selectively modifying the 2DEG in specific regions of the device.
This involves depleting charges underneath the screening gates and accumulating them underneath the plunger and barrier gates (often collectively called \textit{finger gates}) such that the current flows between the screening gates along the desired path, as seen in the first row of Fig.~\ref{fig:physics_steps}(d). 

Determination of voltages at which 1D channels are properly formed is typically done iteratively, by sweeping two screening gates against each other in a 2D measurement while holding the corresponding finger gates constant. 
The formation of the channel manifests itself as a triangular-sloped region (the so-called \textit{triangle plots})~\cite{McJunkin21-PhD}.
If the 2DEG under the finger gates along the channel is not sufficiently accumulated, this region will fail to form. 
Multiple 2D scans with incrementally increased finger-gate voltages for each pair of screening gates---a three-dimensional dataset---are often necessary to observe the triangle region across all required channels.

\begin{figure}[t]
    \centering
    \includegraphics[width=\columnwidth]{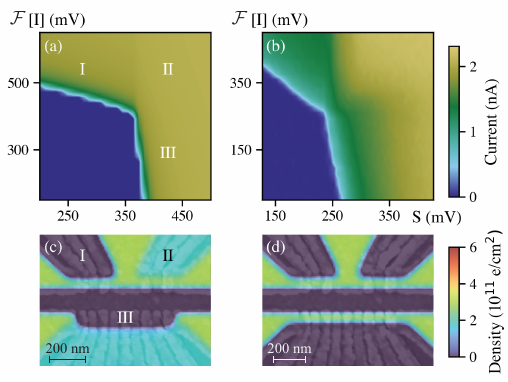}
    \caption{Interplay between screening and finger gates during the single channel formation.
    (a) Current continuity simulation and (b) an example of an experimentally acquired $I$-$V$ response landscape in the screening gate vs finger gate space.
    (c) Schr\"odinger-Poisson simulated example showing three different operation modes, with each current channel programmed to be at a different operation mode: 
    I, a proper 1D channel;
    II, an improper 2D channel under the screening and finger gates;
    III, an improper 2D channel under the screening gate.
    (d) Schr\"odinger-Poisson simulated example of the proper channel formation operation point for the device.
    Details about simulations are given in Supplemental Material~\cite{supp}.
    }
    \label{fig:channel-formation}
\end{figure}

BATIS uses an alternative approach to simultaneously initialize multiple 1D channels. 
Rather than sweeping screening gates against each other, BATIS performs a 2D measurement of a screening gate against all the finger gates located between that gate and the opposing screening gate used to define the 1D channel (typically the central screening gate). 
Thus, this method efficiently captures the behavior of all channels in a single 2D dataset. 

A simulation of current as a function of gate voltages is shown in Fig.~\ref{fig:channel-formation}(a), and in Fig.~\ref{fig:channel-formation}(b) a corresponding example of experimental data is shown.
Notably, there are four unique regions in the landscape in Fig.~\ref{fig:channel-formation}(a): a no-current region at the bottom left, a finger-gate-dominated region at the top left (indicated by ``I''), a screening-gate-dominated region at the bottom right (indicated with ``III''), and a region of combined effect from both the finger gates and the screening gates (indicated by ``II'').
These regions all correspond to different operating modes of the device, with region I showing the proper accumulation of a 1D channel that is the required as a precursor to proper bootstrapping of the device; region II, in contrast, corresponds to an improper 2D channel formed under the screening and finger gates; and region III corresponds to an undesired 2D channel under the screening gate. 
All three of these regions are indicated in the simulation results shown in Fig.~\ref{fig:channel-formation}(c).
If the finger-gate-dominated region does not form, as shown in the right panel in the second row in Fig.~\ref{fig:physics_steps}(d), BATIS initiates illumination with finger gate voltages set more negative than for the most recent illumination.

When the finger-gate-dominated region forms properly, BATIS automatically selects an operating point in region I, near the boundary between the current region and the no-current region. 
Figure~\ref{fig:channel-formation}(d) shows a Schr\"odinger-Poisson simulated example with properly set operating points for all current channels.
The 1D-channel-formation technique implemented in BATIS requires a single measurement per pair of screening and finger gates, making it much more efficient than standard approaches.

\subsection{Finger Gate Characterization}
\label{sec:finger_gate_characterization}
Once isolated current channels are formed, finger gate pinch-offs $V_\mathcal{P}$ can be measured.  
Naively, one might hope that as the voltage on a given finger gate is lowered, the pinch-off curves should exhibit a smooth, sigmoidal trend like that shown in the second row in Fig.~\ref{fig:physics_steps}(e).
However, because finger gates are so small, trapped charge (e.g., in the oxide layer) often causes Coulomb blockade effects~\cite{PhysRevLett.65.771} to be visible, so real devices display behavior like that shown in the third row in Fig.~\ref{fig:physics_steps}(e).
In this case, as the gate voltage is made less positive, the curve exhibits characteristic Coulomb blockade oscillations typical of QD devices.
To be robust against this nonideality, BATIS uses thresholding and slope detection during analysis to determine the pinch-off voltage on a point-by-point basis, going in the direction of increasing voltage from the end of the sweep (the reverse of how the signal was measured) to avoid small deviations of background noise, while capturing the first significant turn-on behavior.

\section{BATIS: Experimental Demonstration}
\label{sec:exp_run}
We demonstrate BATIS at 1.3~\si{\kelvin} on a Si/Si$_x$Ge$_{1-x}$ quadruple QD device with 24 gates, nominally identical to the one shown in Fig.~\ref{fig:physical_problem}(a), which has two charge-sensing channels in parallel with a channel to host the four qubit QDs.
The first gate layer contains four screening gates (S$_1$-S$_4$) shown in Fig.~\ref{fig:physical_problem}(a) in yellow.
The next layer consists of five reservoirs (R$_1$-R$_5$) and six plunger gates (P$_1$-P$_6$). 
The reservoirs are shown in blue in Fig.~\ref{fig:physical_problem}(a), while the gates highlighted in red are the plunger gates.
The final layer consists of nine barrier gates (B$_1$-B$_9$) shown in Fig.~\ref{fig:physical_problem}(a) in gray. 

The goal of BATIS is to form three current channels and determine appropriate pinch-off gate voltages on the basis of automated measurements and the gate layout information BATIS learns from the configuration file. 
This proper operation is indicated by black arrows in Fig.~\ref{fig:physical_problem}(b), bringing a grounded but not-characterized device to a fully operational mode.
In what follows, we explain how BATIS navigates the execution of the bootstrapping, testing, and characterization of the quadruple QD device.

\subsection{First leakage test} 
The sample mount holding the quadruple QD device used here contains $40$ connections, of which $29$ are used for instrument channels to control individual gates and Ohmic contacts, and an additional $11$ wired channels are unused for this device. 
Thus, the leakage matrix is $40 \times 40$ to capture all possible leakage states. 
The leakage matrix is symmetric, and thus only the lower triangular part needs to be completed. 

BATIS first checks the $40$ diagonal elements of the leakage matrix.
If the resistance of any connection drops below 25~\si{\mega\ohm}, BATIS declares that the device leaks. 
Moreover, if a diagonal component fails the threshold test, BATIS automatically fills out the rest of the column for diagnostic purposes. 
In the worst case, a total of $820$ resistance measurements are necessary to complete the leakage matrix.

Under fully automatic control, the leakage test needs to be executed only once per cooldown.
Damage following this test is unlikely, as BATIS incorporates built-in safety checks throughout its automated voltage selection process for all measurements to prevent dielectric breakdown between neighboring gates.

\subsection{Global accumulation}
Next, BATIS seeks to determine gate voltages that establish global accumulation.  
This step generates current flow across the device regardless of which channels are connected to each other.
In our demonstration, the algorithm executed three rounds of illumination before the observed global turn-on values were determined to be satisfactory, occurring in the desired range of $200$\si{\milli\volt} to $400$~\si{\milli\volt}.
By design, all gates are grounded during the first two illuminations.
The first illumination is intended to completely reset the device, while the second is intended to verify that all instruments work as intended.
Because BATIS still had not found the gates in the desired range for the third illumination, voltages were applied to all screening and reservoir gates. 
The voltages required to achieve turn-ons in the required range are determined by BATIS following the protocol in Ref.~\cite{Wolfe24-COI}. 
In this case, during illumination, BATIS sets gates associated with channels I$_1$ and I$_2$ to $280$~\si{\milli\volt}, and gates associated with channel I$_3$ to $300$~\si{\milli\volt}.
The central screening gate, S$_2$, was left grounded (i.e., set to 0~\si{\milli\volt}) during illumination.

\begin{figure}[t]
    \centering
    \includegraphics[width=\columnwidth]{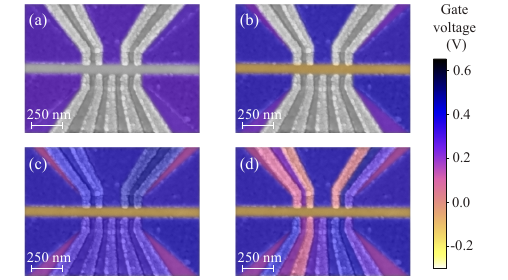}
    \caption{Example gate-voltage configurations determined by BATIS during tuning. 
    The color value indicates the inferred gate voltage, with gates not analyzed at a given stage marked in gray. 
    The color bar applies to all panels.
    (a) Global accumulation. 
    The colors indicate the pinch-off voltages for the reservoir gates.
    (b) Screening and reservoir characterization. 
    The reservoir gates are placed at their operating point, and the exterior screening gates are set to their pinch-off point.
    The central screening gate is set to the guaranteed isolation point.
    (c) Channel accumulation. 
    The finger and screening gates are adjusted to the operating point.
    (d) The device’s final configuration, with the pinch-offs of all the finger gates colored.
    }
    \label{fig:colorful_sems}
\end{figure}

Following the gate-biased illumination, the fourth iteration of the global turn-on measurement was successful.
A summary of all four runs is presented in Table~\ref{tab:global_accum}.
The resulting configuration of the QD device is shown in Fig.~\ref{fig:colorful_sems}(a), where the pinch-off voltages on all the screening gates, excluding S$_2$, and all the reservoir gates are colored on the basis of the learned gate voltage.

\begin{table}[t]
\renewcommand{\arraystretch}{1.1}
\renewcommand{\tabcolsep}{2pt}
\caption{\label{tab:global_accum}
Summary of the four global accumulation runs. 
$V_\mathcal{P}$ represents the detected turn-on voltage, while $V_\mathrm{max}$ is the maximum safe accumulation voltage for a given channel.
BATIS chose to initiate illumination after the first three runs.
The fourth run was satisfactory, and the algorithm proceeded to the next stage.
}
\begin{ruledtabular}
\begin{tabular}{ccS[table-format=3.2]S[table-format=3.2]S[table-format=3.2]S[table-format=3.2]}
 & Voltages & {Run~1} & {Run~2} & {Run~3} & {Run~4} \\
\hline
\multirow{2}{*}{I$_1$} & $V_\mathcal{P}$ (\si{\milli\volt}) &  28.0 & 26.1 & 24.7 & 296.0 \\ 
& $V_\mathrm{max}$ (\si{\milli\volt}) &  81.7 & 78.6 & 75.5 & 349.4 \\ \hline 
\multirow{2}{*}{I$_2$} & $V_\mathcal{P}$ (\si{\milli\volt}) &  28.2 & 20.8 & 23.5 & 308.7 \\ 
& $V_\mathrm{max}$ (\si{\milli\volt}) &  53.6 & 47.3 & 47.3 & 339.2 \\ \hline
\multirow{2}{*}{I$_3$} & $V_\mathcal{P}$ (\si{\milli\volt}) &  7.4 & 9.5 & 2.6 & 317.1 \\ 
& $V_\mathrm{max}$ (\si{\milli\volt}) &  53.6 & 53.6 & 50.4 & 349.4 \\ 
\end{tabular}
\end{ruledtabular}
\end{table}

\subsection{Accumulated Leakage test} 
With the global accumulation confirmed, BATIS proceeds to the accumulated leakage test.
A second leakage matrix is measured, following the same procedure as during the first leakage test.
Like previously, BATIS finds that all resistances exceed 25~\si{\mega\ohm}.
Since none of the connections leaked, only $35$ measurements following the matrix diagonal were taken.

\subsection{Screening and reservoir characterization}
\begin{figure*}[ht!]
    \centering
    \includegraphics[]{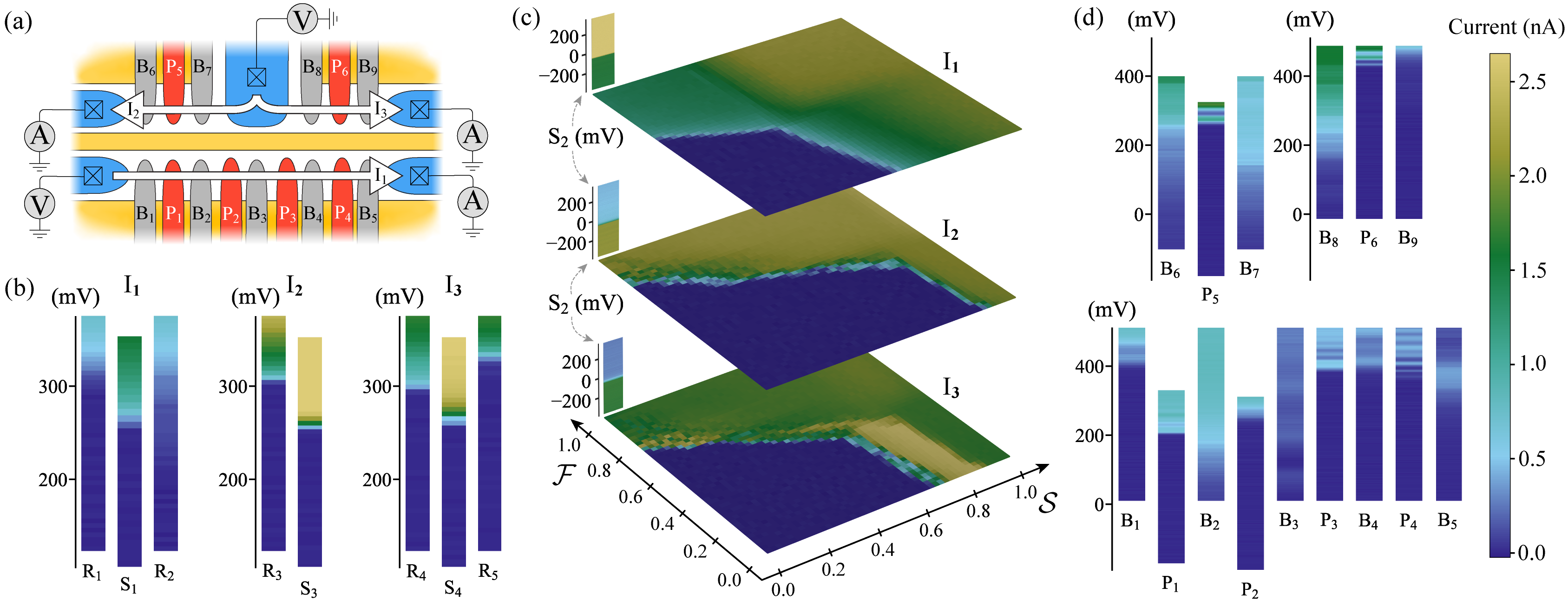}
    \caption{Successful BATIS tune-up on a three-channel, six-QD device.
    (a) Schematic circuit diagram of the QD device shown in Fig.~\ref{fig:physical_problem}(a) depicting how BATIS understands the device.
    Numbered current paths are shown, as are the source voltage and current amplifier connections to Ohmic contacts.
    This information is entered into the configuration file.
    (b) Pinch-off data for all reservoir and external screening gates grouped by measurement channel. 
    The acquired signal for gate S$_1$ extends over twice the voltage range of all other measurements. 
    However, as the low-voltage data far from the transition region exhibit no strong features, the plotted voltage range is truncated for visualization purposes.
    (c) Results of the channel accumulation measurements for each current channel. 
    The $\mathcal{F}$ axis represents the normalized voltage range applied to all of the finger gates, and the $\mathcal{S}$ axis is the normalized voltage range applied to the screening gates bordering a given channel. 
    The finger gates for channel I$_1$ in (a) are swept from $-104$~\si{\milli\volt} to $724$~\si{\milli\volt}, the finger gates for channel I$_2$ in (a) are swept from $-91$~\si{\milli\volt} to $727$~\si{\milli\volt}, and the finger gates for channel I$_3$ are swept from $-83$~\si{\milli\volt} to $729$~\si{\milli\volt}.
    The screening gates are swept from $127$~\si{\milli\volt} to $427$~\si{\milli\volt}, from $-18$~\si{\milli\volt} to $282$~\si{\milli\volt}, and from $50$~\si{\milli\volt} to $350$~\si{\milli\volt} for channels I$_1$, I$_2$, and I$_3$, respectively.
    The S$_2$ pinch-off curves are included for each channel.
    (d) Pinch-off data for all finger gates. 
    BATIS has adjusted the trapped charge under the gates via biased illumination so that all pinch-off voltages fall within a reasonable range, in this case between $-200$~\si{\milli\volt} and $500$~\si{\milli\volt}.
    The color bar on the right corresponds to all plots in (b)--(d).}
    \label{fig:measured_data}
\end{figure*}

The next step in BATIS involves characterization of the screening and reservoir gates, which is important for dividing the induced 2DEG into three current channels. 
The results of these characterizations for all gates tested are presented in Fig.~\ref{fig:measured_data}(b).
Gate S$_2$ borders all of the current channels and thus is not included here as it needs to be measured simultaneously on channels I$_1$, I$_2$, and I$_3$.
The data for the S$_2$ pinch-off can be found in Fig.~\ref{fig:measured_data}(c).

The voltage values are quite consistent between the sets of reservoir and screening gates intended to form the desired current channels, a feature that arises in part from the gate-biased illumination described above~\cite{Wolfe24-COI}.
We note that for gate S$_1$, BATIS acquired the signal over a $600$~\si{\milli\volt} range, i.e., twice as long as for all other screening gates.
This is due to poor fit scores obtained with the reduced-voltage scan range, which prompted BATIS to perform additional measurements with an expanded sweep range to ensure sufficient confidence in the derived pinch-off value.
Since the long tail of the signal has no features, the signal shown in Fig.~\ref{fig:measured_data}(b) is truncated to be consistent with the remaining measurements for visualization purposes.

A summary of the pinch-off voltages and the operating points selected by BATIS is presented in Table~\ref{tab:r_and_s}.
A device in this configuration prepared for the channel accumulation stage is shown in Fig.~\ref{fig:colorful_sems}(b).

\begin{table}[t]
\renewcommand{\arraystretch}{1.1}
\renewcommand{\tabcolsep}{2pt}
\caption{\label{tab:r_and_s}
Summary of the derived voltages for screening and reservoir gates determined during their characterization stage by BATIS.
The pinch-off voltages, $V_\mathcal{P}$, the guaranteed isolation voltages $V_\mathcal{I}$ for screening gates, and the operating points $V_\mathcal{O}$ for reservoirs, are presented.
The gates are organized top-down in the order they were measured.
}
\begin{ruledtabular}
\begin{tabular}{cS[table-format=3.2]S[table-format=3.2]S[table-format=3.2]}
{Gate} & {$V_\mathcal{P}$ (\si{\milli\volt})} & {$V_\mathcal{I}$ (\si{\milli\volt})} & {$V_\mathcal{O}$ (\si{\milli\volt})} \\
\hline
S$_2$ & -54.3 & -176.2 & -103.0 \\ \hline 
S$_1$ & 275.6 & 107.5 & $\cdots$ \\ 
R$_1$ & 291.2 & $\cdots$  & 382.2 \\ 
R$_2$ & 273.4 & $\cdots$  & 390.7 \\ \hline
S$_3$ & 252.4 & 192.5 & $\cdots$ \\ 
R$_3$ & 288.1 & $\cdots$  & 382.5 \\ 
R$_4$ & 273.9 & $\cdots$  & 384.1 \\ \hline
S$_4$ & 269.6 & 180.8 & $\cdots$  \\
R$_4$ & 278.1 & $\cdots$  & 379.3 \\
R$_5$ & 314.2 & $\cdots$  & 370.1 \\
\end{tabular}
\end{ruledtabular}
\end{table}

\subsection{Current channel accumulation}
With all screening and reservoir gates characterized and candidate operating voltages determined, BATIS proceeds to form the 1D current channels.
As discussed in Sec.~\ref{sec:channel_accumulation}, the algorithm for 1D-current-channel formation implemented in BATIS does not follow the typical, triangle-plot-based strategy.
Rather, it sweeps each of the three outer screening gates, i.e., S$_1$, S$_3$, and S$_4$, against all finger gates relevant to the given 1D current channel.
The traditional triangle plots are along a different set of axes (e.g., S$_2$ and S$_1$, instead of all finger gates and S$_1$) than those shown in Fig.~\ref{fig:measured_data}(c).
The advantage of measurements implemented in BATIS is faster convergence to the correct operating point for three well-formed and independent current channels.

The automatic analysis of the regions and their respective boundaries is enabled by fitting a synthetic model to the experimental data, as explained in Ref.~\cite{Schug24-EML}. 
Additional criteria based on the local properties (e.g., estimated resistance and derivatives of the measurement) are used to optimize the selection of the final operating point $V_\mathcal{O}^{I_\ell}$ for $\ell=1,2,3$, among the numerous candidates in the operating region.

If BATIS was unable to select an operating point after such a measurement, the next action would be to adjust all gates using a 400~\si{\milli\volt} illumination.
This adjustment can be found in the configuration file.
After such an illumination, the entire device response can change because of crosstalk and the interdependence of the current paths.
Thus, BATIS resumes autotuning from the global accumulation step.

The quadruple QD device requires three current channels to be formed, as indicated by the white arrows in Fig.~\ref{fig:measured_data}(a).
In this demonstration, BATIS identified three different operating points, one for each 1D current channel, on the basis of the data shown in Fig.~\ref{fig:measured_data}(c).
A summary of the configuration of the screening and finger gates for each operating point can be found in Table~\ref{tab:1d_channel}, and a collapsed visualization of these gate voltages is shown in Fig.~\ref{fig:colorful_sems}(c).

The accumulation plots in Fig.~\ref{fig:measured_data}(c) reveal a dip in the current exiting channel $I_3$ when the current starts to flow in channel $I_2$, shown by the dark green shape imposed on top of the signal in channel $I_3$.
This is due to a coupling between those two channels resulting from a shared Ohmic connection to the current sources.
Such a dip in the charge sensor readout is often called a ``shadow.''
Shadows increase the difficulty of detecting and forming isolated current channels, yet BATIS can handle this situation.
In contrast, the accumulation plot for channel I$_1$ shows much less correlation with the current flow in the two charge sensor channels, although some crosstalk still arises from gate-to-2DEG cross-capacitances.

\subsection{Finger gate characterization} 
The last stage of BATIS involves determining the pinch-off voltages for the finger gates.
One of the problems with analyzing pinch-off data is the presence of spurious Coulomb blockade oscillations, as discussed above and shown in the third row in Fig.~\ref{fig:physics_steps}(e).
Such oscillations complicate fit-based analysis. 
Instead, BATIS uses thresholding and slope detection to determine the pinch-off voltage, searching in the increasing voltage direction (the reverse of how the signal was measured) to capture the first significant turn-on above the noise.

\begin{table}[b]
\renewcommand{\arraystretch}{1.1}
\renewcommand{\tabcolsep}{2pt}
\caption{\label{tab:1d_channel}
Screening and finger gate configuration summary for the three operating points identified by BATIS.
Columns 2--4 provide the screening and finger gate operational configuration, $V^{I_\ell}_\mathcal{O}$ with $\ell=1,2,3$, for each of the operating points identified by BATIS to form channels I$_1$, I$_2$, and I$_3$, respectively.
The finger gates are grouped on the basis of the channel they belong to as $\mathcal{F}[\mathrm{I}_1]$, $\mathcal{F}[\mathrm{I}_2]$, and $\mathcal{F}[\mathrm{I}_3]$.
}
\begin{ruledtabular}
\begin{tabular}{cS[table-format=3.2]S[table-format=3.2]S[table-format=3.2]}
{Gates} & {$V^{I_1}_\mathcal{O}$ (\si{\milli\volt})} & {$V^{I_2}_\mathcal{O}$ (\si{\milli\volt})} & {$V^{I_3}_\mathcal{O}$ (\si{\milli\volt})} \\
\hline
S$_1$ & 189.2 & 276.8 & 228.8 \\ 
$\mathcal{F}[\mathrm{I}_1]$ & 356.6 & 364.3 & 369.5 \\ \hline
S$_3$ & 42.4 & 131.7 & 84.4 \\ 
$\mathcal{F}[\mathrm{I}_2]$ & 389.8 & 397.0 & 401.9 \\ \hline
S$_4$ & 109.6 & 199.6 & 153.3 \\
$\mathcal{F}[\mathrm{I}_3]$ & 456.0 & 462.5 & 466.9 \\
\end{tabular}
\end{ruledtabular}
\end{table}

\begin{table}[t]
\renewcommand{\arraystretch}{1.1}
\renewcommand{\tabcolsep}{2pt}
\caption{\label{tab:finger_final}
Final pinch-off values, as determined by BATIS, for all the finger gates.
The three main columns are associated with a set of finger gates belonging to a particular current channel on the device $\mathcal{F}[\mathrm{I}_\ell]$, with $\ell=1,2,3$.
Each of the main columns consists of two subcolumns, with the left subcolumn listing the finger gates belonging to the selected channel, and the right subcolumn giving the pinch-off voltage $V_\mathcal{P}$ determined by BATIS.
}
\begin{ruledtabular}
\begin{tabular}{cS[table-format=3.2]cS[table-format=3.2]cS[table-format=3.2]}
\multicolumn{2}{c}{$\mathcal{F}[\mathrm{I}_1]$} & \multicolumn{2}{c}{$\mathcal{F}[\mathrm{I}_2]$} & \multicolumn{2}{c}{$\mathcal{F}[\mathrm{I}_3]$} \\
Gate & $V_\mathcal{P}$~{(\si{\milli\volt})} & 
Gate & $V_\mathcal{P}$~{(\si{\milli\volt})} & 
Gate & $V_\mathcal{P}$~{(\si{\milli\volt})} \\
\hline
B$_1$ & 406.1 & B$_6$ & 84.0 & B$_8$ & 32.2\\
P$_1$ & 203.0 & P$_5$ & 252.7 & P$_6$ & 424.3 \\
B$_2$ & 87.7 & B$_7$ & 14.5 &  B$_9$ & 339.7 \\
P$_2$ & 303.7 & & & & \\
B$_3$ & 318.2 & & & & \\
P$_3$ & 387.4 & & & & \\
B$_4$ & 395.7 & & & & \\
P$_4$ & 399.6 & & & & \\
B$_5$ & 321.8 & & & & \\
\end{tabular}
\end{ruledtabular}
\end{table}

This pinch-off voltage determination is performed individually for each finger gate of the QD device.
The voltage range for each measurement is always the same and predefined in the configuration file.
However, because of the Coulomb blockade oscillations, the automated search for the starting voltage for some gates is more challenging than for others, as is visible in the data shown in Fig.~\ref{fig:measured_data}(d), and the added complexity of the curves informed the algorithm described above and implemented here.
The resulting pinch-off values of all these differing curves for all finger gates are presented in Table~\ref{tab:finger_final}.
This configuration is used in the final colorized scanning electron microscopy image for the ending state of the algorithm shown in Fig.~\ref{fig:colorful_sems}(d). 
This voltage configuration describes the proper operating point in high-dimensional voltage space.
BATIS has taken the device from an unknown initial state to a set of properly defined 1D channels, with confirmed fingergate operation and a full compilation of pinch-off voltages.
At this point, autotuners designed for 1D devices can be used on each individual channel, if desired~\cite{Zwolak21-AAQ}.

\section{Summary and outlook}
\label{sec:conclusion}
In this paper, we demonstrated an autonomous system for bootstrapping, automated testing, and characterization of gate-defined, accumulation-mode, undoped QD devices. 
This method, which we call BATIS, leverages knowledge from experiments and physical intuition---grounded in fundamental device physics---to electronically control the navigation of the quantum device landscape.
By being rooted in heuristics and physical principles, the non-machine-learning implementation of BATIS provides a robust and interpretable baseline for developing and benchmarking future model-free strategies.

The prior knowledge about the QD device required to initialize BATIS is limited to the heterostructure type, the device geometry, the Ohmic biases, and the device safety limits.
The method proceeds in several steps, checking device functionality at each step.  
If irrecoverable device flaws are discovered, BATIS reports those issues and halts. 
Importantly, BATIS can recover from other nonidealities, such as undesirable amounts of trapped charge arising from fabrication or accidents, by using illumination under gate bias to modify \textit{in situ} the trapped charge and thus the device performance.
This capability is important for resolving issues such as charge disorder or incomplete turn-on, which may prevent functional operation in otherwise operational devices.

While much of the automated approach relies on existing experimental procedures, an innovation in BATIS is its method for 1D channel formation: rather than sweeping pairs of screening gates, BATIS performs a single 2D voltage scan between one screening gate and all intermediate finger gates. 
This strategy reduces the required measurement overhead to a single scan, resulting in substantial efficiency gains compared with conventional approaches. 

BATIS is built for scalability and extensibility. 
Its flexible architecture supports characterization and initialization of any accumulation-mode linear QD array and can be adapted to other QD platforms with minimal modification. 
For example, BATIS can replicate tuning routines developed for nanowires or GaAs/Al$x$Ga${1-x}$As devices by simply skipping steps unnecessary for those architectures. 
In contrast, it is quite unlikely that any previously proposed algorithms could be easily adapted to the Si/Si$_x$Ge$_{1-x}$ quadruple QD device considered in this work.

Importantly, the single-button-press testing and characterization process is demonstrated at $1.3$~\si{\kelvin}, eliminating the need for a full millikelvin cooldown for the device initialization. 
This result is particularly important as work advances toward ``hot,'' large-scale spin-based quantum processors.
BATIS can also serve as a screening tool before cooling a dilution refrigerator to reach millikelvin temperatures, if cold operation is desired.

Previous studies have demonstrated device-agnostic algorithms for certain subsets of QD-device-tuning tasks, such as setting QD topologies~\cite{Ziegler22-TRA, Zubchenko24-ABQ} or charge-state classifications~\cite{Durrer19-ATQ, Ziegler22-TAR}.
These algorithms presuppose that device characterization and foundational tuning steps---such as determining gate functionality, defining the QD channel(s), or locating regions of interest for the coarse-tuning regime---have already been performed.
BATIS helps to fill this gap, enabling bootstrapping of gate-defined devices in which current paths are entirely defined by gate voltages whose values cannot be known in advance.

\begin{acknowledgments}
We acknowledge Danielle Middlebrooks for helping with the automation.
We acknowledge HRL Laboratories, LLC, for support and L.F. Edge for providing one of the Si/Si$_x$Ge$_{1-x}$ heterostructures used in this work. 
This work was supported in part by ARO Grants No. W911NF-17-1-0274, No. W911NF-23-1-0110, and No. W911NF-24-2-0043.
The authors gratefully acknowledge the use of facilities and instrumentation in the University of Wisconsin-Madison Wisconsin Center for Nanoscale Technology, which is partially supported by the Wisconsin Materials Research Science and Engineering Center (NSF Grant No. DMR-2309000) and the University of Wisconsin-Madison. 
The views and conclusions contained in this paper are those of the authors and should not be interpreted as representing the official policies, either expressed or implied, of the U.S. Government. 
The U.S. Government is authorized to reproduce and distribute reprints for Government purposes notwithstanding any copyright noted herein. 
Any mention of commercial products is for information only; it does not imply recommendation or endorsement by NIST.
\end{acknowledgments}

\section*{Data availability}
The data that support the findings of this article are openly available~\cite{qf-bootstrap}.

%


\clearpage
\onecolumngrid

\includepdf[pages=1,fitpaper=true,pagecommand={}]{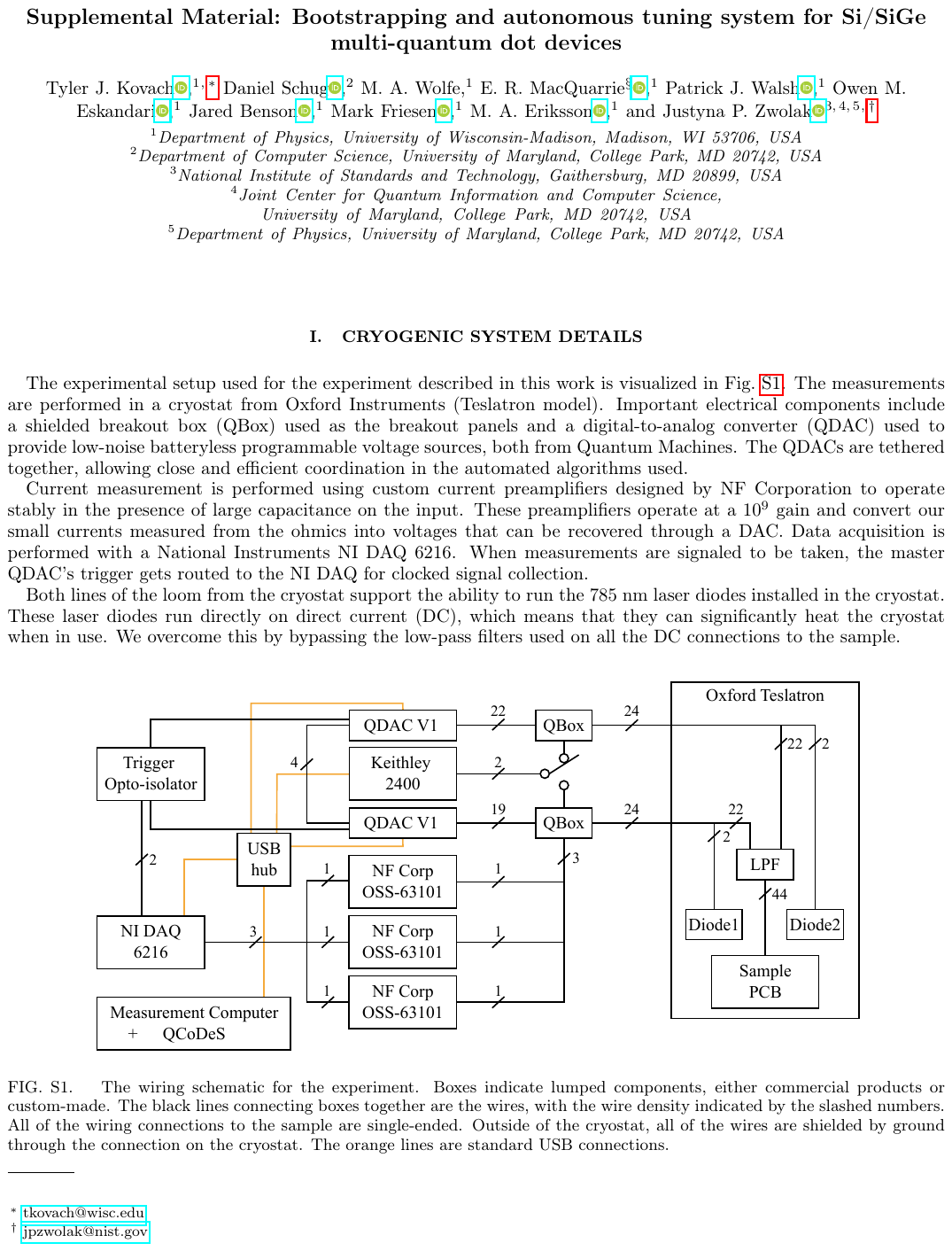}
\includepdf[pages=2,fitpaper=true,pagecommand={}]{supplement.pdf}
\includepdf[pages=3,fitpaper=true,pagecommand={}]{supplement.pdf}

\end{document}